\documentclass[prl,twocolumn,showpacs,preprintnumbers,amsmath,amssymb,superscriptaddress]{revtex4}

\usepackage{graphicx} 
\usepackage{color} 
\usepackage{bbm, bm,fancybox}
\usepackage{euscript,eufrak,oldgerm}
\newcommand{\ave}[1]{\left\langle #1 \right\rangle} 

\newcommand{\ie}{{\it{i.e.~}}}
\newcommand{\etal}{{\it{et al.}}}

\begin{document}

\title{Quantum Non-locality and Partial Transposition for Continuous-Variable Systems}

\author{Alejo Salles}
\email{salles@if.ufrj.br} \affiliation{Instituto de F\'{\i}sica, Universidade
  Federal do Rio de Janeiro, 
  Rio de Janeiro, RJ
  21941-972, Brazil} \affiliation{Albert-Ludwigs-Universit\"at Freiburg, Physikalisches Institut, 
  D-79104 Freiburg, Germany}

\author{Daniel Cavalcanti}
\email{daniel.cavalcanti@icfo.es}
\affiliation{ICFO-Institut de
Ciencies Fotoniques, Mediterranean Technology Park, 08860
Castelldefels (Barcelona), Spain}

\author{Antonio Ac\'in}
\email{antonio.acin@icfo.es}
\affiliation{ICFO-Institut de
Ciencies Fotoniques, Mediterranean Technology Park, 08860
Castelldefels (Barcelona), Spain}
\affiliation{ICREA-Instituci\'o
Catalana de Recerca i Estudis Avan\c cats, Lluis Companys 23,
08010 Barcelona, Spain}


%
%

\begin{abstract}
A continuous-variable Bell inequality, valid for an arbitrary
number of observers measuring observables with an arbitrary number
of outcomes, was recently introduced in [Cavalcanti \emph{et al.},
Phys. Rev. Lett. {\bf 99}, 210405 (2007)]. We prove that any
$n$-mode quantum state violating this inequality with quadrature
measurements necessarily has a negative partial transposition. Our
results thus establish the first link between nonlocality and
bound entanglement for continuous-variable systems.
\end{abstract}

\maketitle

%
%
During most of the history of quantum mechanics, the concepts of
entanglement and non-locality were considered as a single feature
of the theory. It was based on the discussion of non-locality started by
Einstein, Podolsky and Rosen~\cite{epr} that Schr\"odinger
stressed the importance of entanglement in the understanding of
quantum systems~\cite{schrodinger}. Later, Bell derived
experimentally testable conditions, known as Bell inequalities, to
verify the non-local character of entanglement~\cite{bell}.

It was only with the recent advent of quantum information science
that the relation between these two concepts started to be
considered in depth. On one hand, we need
entanglement for a state to be nonlocal, where by non-locality we
understand incompatibility with local-hidden-variable (LHV) models
\cite{gisin}. But, on the other hand, we know that some entangled
states admit a LHV model~\cite{werner}. The situation is even
richer due to the fact that there exist other meaningful scenarios
where sequences of measurements \cite{sequences} or
activation-like processes \cite{MLD} allow detecting the hidden
non-locality of quantum states. More in general, the relation
between these concepts is still not fully understood. Clarifying
this relation is highly desirable, for it would lead us to
ultimately grasp the very essence of quantum mechanics.

One way to tackle this problem is by studying the relation between
non-locality and other concepts regularly related to entanglement.
In this direction, Peres conjectured~\cite{peres2} that any state
having a positive partial transposition (PPT) should admit a LHV
model, or, equivalently, any state violating a Bell inequality
should have a negative partial transposition (NPT). Partial
transposition has been proven one of the most useful tools in the
study of entanglement. As shown by Peres~\cite{peres}, any NPT
state is necessarily entangled. However, positivity of the partial
transposition (PPT) represents a necessary, but not sufficient,
condition for a state to be separable. Indeed, partial
transposition is just the simplest example of positive maps,
linear maps that are useful for the detection of mixed-state
entanglement~\cite{horodecki}.
A second fundamental result on the connection between partial
transposition and entanglement was to notice that all PPT states
are non-distillable \cite{HorodeckiBound}. In other words, if an
entangled state is PPT, it is impossible to extract pure-state
entanglement out of it by local operations assisted by classical
communication (LOCC), even if the parties are allowed to perform
operations on many copies of the state. In this way, PPT
entanglement was regarded, for a long time, as a useless kind of
quantum correlations \cite{boundcrypto}.

Proving (or disproving) Peres' conjecture in full generality
represents one of the open challenges in quantum information
theory. The proof of the conjecture has up to now been achieved
only for some particular cases; if we label the nonlocality
scenario as is customary by the numbers $(n, m, o)$, meaning that
$n$ parties can choose between $m$ measurement settings of $o$
outcomes each, the most general proof obtained so far is for
correlation Bell inequalities in the $(n, 2, 2)$
case~\cite{werner2,dist}. To our knowledge, there is no result for
larger number of settings and/or outcomes and, in particular, for
the relevant case of continuous-variable (CV) systems. These
systems are very promising for loophole-free Bell violations, due
to the high efficiency achieved in homodyne
detection~\cite{loophole-free}.

Unfortunately, there have been few results so far on Bell
inequalities for CV systems~\cite{cv-inequalities}. 
Recently, Cavalcanti, Foster, Reid and Drummond (CFRD) introduced a very general Bell inequality for the $(n,2,o)$ scenario with arbitrary $n$ and $o$, which works in particular when $o\rightarrow\infty$, the CV case.
We make use of the Shchukin and Vogel (SV) NPT criterion~\cite{shchukin}
to show that the CFRD inequality with two arbitrary quadratures as
settings on each site is not violated for multipartite PPT states,
thus proving Peres' conjecture in this relevant scenario. This is
the first result on the
connection between partial transposition and Bell 
inequalities for CV systems, and takes us a step further in
understanding the relation between entanglement and non-locality.
After proving this result, we discuss on the practical
applicability of the CFRD inequality, showing that no two-
mode quantum state can violate 
it when performing two homodyne measurements on each site.

\emph{CFRD Bell inequality.--}
In Ref.~\cite{cavalcanti}, the authors use the fact that the variance of any function of random variables must necessarily be positive to get general Bell inequalities.
By choosing functions of local observables one can find
discrepancies between the quantum and the classical predictions
just using the fact that in the quantum case these observables are
given by Hermitian operators (usually satisfying non-trivial
commutation relations), while in an LHV scenario the observables
are just numbers, given a priori by the hidden variables (and
obviously commuting with each other). Interestingly, this idea can
lead to strong Bell inequalities as 
the the well-known Mermin, Ardehali, Belinskii and Klyshko inequality~\cite{mabk}.
More importantly for the present discussion, the CFRD approach
works for unbounded observables as well, leading to Bell
inequalities for CV systems.

Consider a complex function $C_n$ of the local real
observables $\{X_k, Y_k\}$,  where $k$ labels the different parties,
defined as:
\begin{equation}\label{Cn}
C_n = \tilde{X}_n+ i \tilde{Y}_n = \prod_{k=1}^{n} (X_k + i Y_k),
\end{equation}
Applying the positivity of the variance of both its real ($
\tilde{X}_n$) and imaginary ($\tilde{Y}_n$) part, and assuming LHV
(\ie setting commutators to zero) we obtain~\cite{cavalcanti}:
\begin{equation}
\label{InequalityTwoObservables} \langle \tilde{X}_n\rangle^2+
\langle \tilde{Y}_n\rangle^2 \leq
\left\langle\prod_{k=1}^n(X^2_k+Y^2_k)\right\rangle.
\end{equation}
This inequality must be satisfied by LHV models for any set of
observables $\{X_k, Y_k\}$, regardless of their spectrum. In
particular, it applies to CV observables.

Consider, for each site, two arbitrary quadratures defined in
terms of the annihilation (creation) operators $\hat{a}_k$
($\hat{a}^\dagger_k$) as:
\begin{equation}
\label{NonOrthogonalQuadraturesDefinition}
\begin{split}
\hat{X}_k&=\hat{a}_k e^{-i\theta_k}+\hat{a}^{\dagger}_k e^{i\theta_k},\\
\hat{Y}_k&=\hat{a}_k
e^{-i(\theta_k+\delta_k+s_k\pi/2)}+\hat{a}^{\dagger}_k
e^{i(\theta_k+\delta_k+s_k\pi/2)},
\end{split}
\end{equation}
where $-\pi/2<\delta_k<\pi/2$ quantifies the departure from
orthogonality, $s_k = \pm1$, and $[a_k,a_k^\dag]=1$. With these
parameters all possible choices of angles are covered, noting that
$\delta_k=-\pi/2,\pi/2$ corresponds to measuring only 
one quadrature. Note that the measurement scenario used in Ref.
\cite{cavalcanti} is a particular case of
\eqref{NonOrthogonalQuadraturesDefinition} corresponding to
$\delta_k=0$. We define new operators $\hat{b}_k$ and
$\hat{b}^\dagger_k$ as:
\begin{equation}
\hat{b}_k=\frac{(\hat{X}_k+e^{is_k\pi/2}\hat{Y}_k)e^{i\theta_k}}{2\sqrt{\cos\delta_k}};
\hat{b}^\dagger_k=\frac{(\hat{X}_k+e^{-is_k\pi/2}\hat{Y}_k)e^{-i\theta_k}}{2\sqrt{\cos\delta_k}}.
\end{equation}
These operators satisfy the usual commutation relation
$[\hat{b}_k,\hat{b}^\dagger_k]=1$, independently of the $s_k$.
Inverting these equations we get:
\begin{equation}
\begin{split}
\label{NonorthogonalQuadratures}
\hat{X}_k&=\sqrt{\cos\delta_k}\left(\hat{b}_k
e^{-i\theta_k}+\hat{b}^{\dagger}_k e^{i\theta_k}\right),\\
\hat{Y}_k&=\sqrt{\cos\delta_k}\left(\hat{b}_k  e^{-i(\theta_k+s_k
\pi/2)}+\hat{b}^{\dagger}_k e^{i(\theta_k+s_k \pi/2)}\right).
\end{split}
\end{equation}

Plugging these operators in \eqref{InequalityTwoObservables}  we
arrive at the CFRD inequality for \emph{arbitrary}
quadratures:
\begin{equation}
\label{RenormalizedNonOrthogonalQuadratureInequality}
\left|\ave{\prod_{k}\hat{B}_k(s_k)}\right|^2\leq\frac{1}{\prod_{k}\cos\delta_k}\ave{\prod_k\left(\cos\delta_k\hat{b}_k^{\dagger}\hat{b}_k+\frac{1}{2}\right)},
\end{equation}
 with $\hat{B}_k(1)=\hat{b}_k$ and $\hat{B}_k(-1)=\hat{b}^{\dagger}_k$.
In what follows, we show that all states violating this inequality
must be NPT according to some bipartition. In order to prove this,
we need to recall Shchukin and Vogel's (SV)
criterion~\cite{shchukin}.

\emph{SV criterion.--} A necessary and sufficient condition for the
positivity of the partial transposition of a CV state, given in terms
of matrices of moments, was introduced and further generalized to
the multipartite case in~Refs.~\cite{shchukin}. 
When dealing with many parties, one must analyze the positivity of
the partial transposition for the different partitions of the
system into two groups.  
We say that a state is PPT when it is
PPT according to \emph{all} bipartitions. Let us introduce the SV
criterion for the multipartite scenario.

When considering the partial transposition of a quantum state $\rho$ with respect to a given
bipartition of the system, we label by $I$ the set of parties that
we choose to transpose, which also defines the corresponding
bipartition. We construct a matrix of moments $M^I$ whose elements
are given by the expectation values:
\begin{equation}
\label{MatrixElements} M^I_{st}=\ave{\prod_{i\in
I}\hat{b}_i^{\dagger q_i}\hat{b}_i^{p_i}\hat{b}_i^{\dagger k_i}
\hat{b}_i^{l_i}\prod_{i\in \bar{I}}\hat{b}_i^{\dagger
l_i}\hat{b}_i^{k_i}\hat{b}_i^{\dagger p_i}\hat{b}_i^{q_i}},
\end{equation}
where ${\bf k}=(k_1,\ldots,k_n)$ and ${\bf l}=(l_1,\ldots,l_n)$
correspond to row index $s$, and ${\bf p}=(p_1,\ldots,p_n)$ and
${\bf q}=(q_1,\ldots,q_n)$ correspond to column index $t$, with
some prescribed ordering that is not relevant for our purposes (see~\cite{shchukin} for details); and $\bar{I}$
denotes the complement of $I$, that is, those parties which we
choose \emph{not} to transpose. We stress that, for fixed row and
column indices, the ordering of the operators entering the
corresponding matrix element will depend on the bipartition $I$.

Shchukin and Vogel's criterion says that, for a state to be PPT
according to bipartition $I$, all principal minors of $M^I$ should
be nonnegative~\cite{ppalminors}. So, for a state to be PPT according to \emph{all} bipartitions, all
principal minors of \emph{all} matrices $M^I$ must be nonnegative,
for all nontrivial bipartitions $I$. By nontrivial
bipartitions we mean that we exclude the bipartition labeled by
$I=\emptyset$, as well as that labeled by $I=\cal{N}$, the entire
set, both corresponding to no transposition at all. In these
cases, the criterion speaks about the positivity of the
state itself, instead of its partial transposition.

\emph{Nonlocality implies NPT.--} We are now in position of proving
that any state violating the generalized CFRD
inequality~\eqref{RenormalizedNonOrthogonalQuadratureInequality} is necessarily NPT. We
begin by expanding the products in the  RHS of
inequality~\eqref{RenormalizedNonOrthogonalQuadratureInequality} as follows:
\begin{widetext}
\begin{eqnarray}\label{NonOrthogonalProductExpand}
&&\frac{1}{\prod_k \cos\delta_{k}}\ave{\prod_k \left(\cos\delta_k
\hat{N}_k+\frac{1}{2}\right)}=\frac{1}{\prod_k
\cos\delta_{k}}\left(\frac{1}{2^n}+\frac{1}{2^{n-1}}\sum_{i_1=1}^{n}\cos\delta_{i_1}\ave{\hat{N}_{i_1}}+\frac{1}{2^{n-2}}\sum_{i_1=1}^{n}\sum_{i_2>i_1}^{n}\cos\delta_{i_1}\cos\delta_{i_2}\nonumber\right.\\&&\left.\ave{\hat{N}_{i_1}\hat{N}_{i_2}}+\ldots+\frac{1}{2}\sum_{i_1=1}^{n}\sum_{i_2>i_1}^{n}\cdots\sum_{i_{n-1}>i_{n-2}}^n\cos\delta_{i_1}\cos\delta_{i_2}\cdots\cos\delta_{i_{n-1}}\ave{\hat{N}_{i_1}\hat{N}_{i_2}\cdots\hat{N}_{i_{n-1}}}\right)+\ave{\prod_k
\hat{N}_k},
\end{eqnarray}
\end{widetext}
where we made use of the number operators defined as $\hat{N}_k\equiv\hat{b}_k^{\dagger}\hat{b}_k$.
We now take all but
the last term on the RHS of eq.~\eqref{NonOrthogonalProductExpand} and call
their sum $S^2$, so that:
\begin{equation}
\frac{1}{\prod_k \cos\delta_{k}}\ave{\prod_k \left(\cos\delta_k \hat{N}_k+\frac{1}{2}\right)}=S^2+\ave{\prod_k \hat{N}_k}.
\end{equation}
Note that $S^2$ is a nonnegative quantity, since $\-\pi/2<\delta_k<\pi/2$ and the expectation value of a product of number operators
is always nonnegative.
We can rewrite inequality \eqref{RenormalizedNonOrthogonalQuadratureInequality} as:
\begin{equation}
\label{NonorthogonalSimpleInequality}
\ave{\prod_k \hat{N}_k}-\ave{\prod_k \hat{B}_k(s_k)}\ave{\prod_k \hat{B}_k(-s_k)}\geq-S^2.
\end{equation}
The key point in the proof is to realize that, for any choice of the parameters $s_k$, the left-hand side (LHS) of
eq.~\eqref{NonorthogonalSimpleInequality} is just one of the principal minors
of $M^I$, provided we choose the bipartition $I$ appropriately. The
principal minor we will look at is:
\begin{equation}
D^I=\left|\begin{array}{cc}
1&\ave{\prod_k \hat{B}_k(s_k)}\\
\ave{\prod_k \hat{B}_k(-s_k)} &\eta_I
\end{array}\right|,
\end{equation}
where $\eta_I$ depends on the bipartition $I$, and which we want to take the form $\eta_I=\ave{\prod_k \hat{N}_k}$.

Looking at the elements of the matrix of moments $M^I$ given by
eq.~\eqref{MatrixElements}, we note that the indices labeling the
diagonal element that has one creation operator
$\hat{b}_k^\dagger$ and one annihilation operator $\hat{b}_k$ in
normal order are $l_k=1$, $k_k=0$, $p_k=0$ and $q_k=1$. The
corresponding upper right element is in turn labeled, for the $k$
part, by $l_k=0$, $k_k=0$, $p_k=0$ and $q_k=1$. If we have the
choice of setting $s_k=-1$ we want this to correspond to a
creation operator $\hat{b}_k^\dagger$ appearing in this position,
which means that our bipartition must be such that $I$
includes site $k$. Conversely, if we have, for a different $k$,
$s_k=1$, site $k$ should \emph{not} be in $I$.

Hence, if we choose the bipartition as that labeled by $I$
including all sites with setting $s_k=-1$, we get
$\eta_I=\ave{\prod_k \hat{N}_k}$, and thus:
\begin{equation}
D^I=\ave{\prod_k \hat{N}_k}-\ave{\prod_k \hat{B}_k(s_k)}\ave{\prod_k \hat{B}_k(-s_k)}.
\end{equation}
It follows that a violation of
inequality~\eqref{NonorthogonalSimpleInequality} implies that
$D^I<0$, and the violating state must be NPT according to
bipartition $I$, or just NPT, which concludes the proof.

We note that if all $s_k$ are equal to either $1$ or $-1$, this
corresponds respectively to $I=\emptyset$ or $I=\cal{N}$, meaning
no transposition at all. As we mentioned above, in this case the
positivity of the minors speaks no longer about the positivity of
the partial transpose of the state but about the positivity of the
state itself. A violation for this choice of parameters, thus,
would mean that the state is not positive semidefinite, which is
unphysical.

\emph{Applicability of the CFRD inequality.--}
Before concluding, we would like to discuss about the
applicability of the CFRD inequality. In particular, we now show
that in the case of two parties, the CFRD inequality is never
violated for measurements on two quadratures per site. An example
of violation of this inequality corresponding to measurements on
orthogonal quadratures applied to a ten-mode cat-like state was
given in the original reference~\cite{cavalcanti}. There, it was
also shown that the quantum violation of the inequality increases
exponentially with the number of modes~\cite{cavalcanti}. However,
in spite of its elegance and conceptual relevance, at present
there is no feasible scheme \cite{notefeas} producing a violation
of the CFRD inequality. This remains as an interesting open
question.

Let us start by considering systems of two parties with measurements on arbitrary quadratures. Applying the positivity of the variance for the real and imaginary parts of $C_2$ (see \eqref{Cn}) without neglecting the terms containing commutators we get:
\begin{equation}
\label{QuadratureInequalityn=2}
\underbrace{\langle \tilde{X}_2\rangle^2+ \langle \tilde{Y}_2\rangle^2 - \left\langle\prod_{k=1}^2(X^2_k+Y^2_k)\right\rangle}_{\beta_2}\leq
 -\left \langle[ X_1,Y_1][ X_2,Y_2]\right\rangle
\end{equation}
The Bell inequality \eqref{InequalityTwoObservables} follows by setting the right-hand side (RHS) of this inequality to zero, and we are left with $\beta_2\leq0$, since for LHV models all commutators are null. So, in order to have a violation we need to find a state such that $\beta_2>0$. We are going to show that  this never happens with the choice~$\eqref{NonOrthogonalQuadraturesDefinition}$. Indeed, taking the settings of~$\eqref{NonOrthogonalQuadraturesDefinition}$, the RHS of \eqref{QuadratureInequalityn=2} becomes $4s_1 s_2\cos\delta_1\cos\delta_2$, so we have $\beta_2\leq 4s_1 s_2\cos\delta_1\cos\delta_2$.
If the parameters are chosen to be different, \ie $s_1=-s_2=\pm1$,
we have that $\beta_2\leq-4\cos\delta_1\cos\delta_2<0$ for all
quantum states, and then there is no violation in this case. As we
have previously discussed for arbitrary $n$, no violation can take
place for the case in which all the $s_k$ parameters are equal,
$s_1=s_2$.

\emph{Concluding remarks.--} Despite years of effort, little is
known about which entangled states admit a LHV model, that is, can
be simulated using classical correlations. Peres' conjecture
represents one of the most interesting open problems related to
this fundamental question. Our results are the first to provide support for its validity in the CV regime.
It is also the first result beyond the two-setting two-outcome scenario, since all previous
partial proofs of the conjecture were for the case of two
measurements of two outcomes per site. This gives more support to
the belief that the impossibility of distilling entanglement is
intimately linked to the existence of an LHV model for a given
quantum state.

Finally, CV Bell inequalities suitable for practical tests are very desired due to the the high control attained in CV photonic experiments that will allow a loophole-free demonstration. We have shown here that the CFRD can never be violated by two-mode states with quadrature measurements. It is then a relevant open question to construct Bell inequalities suitable for CV systems that can be violated by states consisting of a small number of modes. Future research in this direction could for instance involve the study of CV Bell inequalities involving more measurements per site~\cite{SV}.

%
%

\begin{acknowledgments}
We acknowledge fruitful discussions with S.
Walborn, R. L. de Matos Filho, A. Ferraro, E. Shchukin and W. Vogel.
A.S. and D.C. thank
Andreas Buchleitner's group for their hospitality in Freiburg.
This work was financially supported by CAPES/DAAD, FAPERJ, the Brazilian
Millennium Institute for Quantum Information, the EU projects COMPAS and QAP, the Spanish projects FIS2007-60182 and
Consolider QOIT, and the Generalitat de Catalunya.
\end{acknowledgments}

%
%


\begin{thebibliography}{20}
\bibitem{notefeas}
A feasible scheme could, for instance, consist of (i) a state of a small number of modes whose preparation requires a few non-Gaussian operations and (ii) homodyne measurements.
\end{thebibliography}
\end{document}